\documentclass[5p,times,twocolumn,compress]{elsarticle}

\usepackage{amssymb}
\usepackage{amsmath}
\newcommand{\bol}[1]{\boldsymbol{#1}}
\usepackage{physics}
\usepackage{calc}
\usepackage{mathtools}
\usepackage{enumitem}
\usepackage[pdftex,pdfpagelabels,bookmarks,hyperindex,hyperfigures,colorlinks]{hyperref}

\journal{Future Generation Computer Systems}

\begin{document}

\begin{frontmatter}



\title{Quantum simulation of dissipation for Maxwell equations in  dispersive media}


\author[label1]{Efstratios Koukoutsis\corref{cor1}}
\ead{stkoukoutsis@mail.ntua.gr}
\author[label1]{Kyriakos Hizanidis}
\author[label2]{Abhay K. Ram}
\author[label3]{George Vahala}

\affiliation[label1]{organization={School of Electrical and Computer Engineering, National Technical University of Athens},
            city={Zographou},
            postcode={15780}, 
            country={Greece}}
            \cortext[cor1]{Corresponding author}
\affiliation[label2]{organization={Plasma Science and Fusion Center, Massachusetts Institute of Technology},
            city={Cambridge},
            state={Massachusetts},
            postcode={02139}, 
            country={USA}}
\affiliation[label3]{organization={Department of Physics, William \& Mary},
            city={Williamsburg},
            state={Virginia},
            postcode={23187}, 
            country={USA}}
            
\begin{abstract}
The dissipative character of an electromagnetic medium breaks the unitary evolution structure
that is present in lossless, dispersive optical media. In dispersive media, dissipation appears in the Schr\"odinger representation of classical Maxwell equations as a sparse diagonal operator occupying an $r$-dimensional subspace. A first order Suzuki-Trotter approximation for the evolution operator enables us to isolate the non-unitary operators (associated with dissipation) from the unitary operators (associated with lossless media). The unitary operators can be implemented through qubit lattice algorithm (QLA) on $n$ qubits, based on the discretization and the dimensionality of the pertinent fields. However, the non-unitary-dissipative part poses a challenge both physically and computationally on how it should be implemented on a quantum computer. In this paper, two probabilistic dilation algorithms are considered for handling the dissipative operators. The first algorithm is based on treating the classical dissipation as a linear amplitude damping-type completely positive trace preserving (CPTP) quantum channel where an unspecified environment interacts with the system of interest and produces the non-unitary evolution. Therefore, the combined system-environment is now closed, and must undergo unitary evolution in the dilated space. The unspecified environment can be modeled by just one ancillary qubit, resulting in an implementation scaling of $\textit{O}(2^{n-1}n^2)$ elementary gates for  the total system-environment unitary evolution operator. The second algorithm approximates the non-unitary operators by the Linear Combination of Unitaries (LCU). On exploiting the diagonal structure of the dissipation, we obtain an optimized representation of the  non-unitary part, which requires $\textit{O}(2^{n})$ elementary gates. Applying the LCU method for a simple dielectric medium with homogeneous dissipation rate, the implementation scaling can be further reduced into $\textit{O}[poly(n)]$ basic gates. For the particular case of weak dissipation we show that our proposed post-selective dilation algorithms can efficiently delve into the transient evolution dynamics of dissipative systems by calculating the respective implementation circuit depth.
A connection of our results with the non-linear-in-normalization-only (NINO) quantum channels is also presented.
\end{abstract}




\end{frontmatter}


\section{Introduction}\label{sec:1}

Electromagnetic waves are ubiquitous in natural and artificial environments and play a useful role in a variety of applications ranging from communications to heating of thermonuclear fusion plasma. The propagation of waves through different dielectric and magnetic media is usually studied using computational models, the implementation of which on classical computers is not cost effective either in terms of available resources or in computational run time. However, the possibility that quantum computers could overcome the limitations of classical computing \cite{Arute,Wu} has been the motivation behind research on the application of quantum information science (QIS) to traditionally classical fields. In this direction, Maxwell equations, which  mediate the physics of propagation and scattering of electromagnetic waves in complex media, are appropriate for quantum implementation due to their linear nature and that can be recast into a form that is similar to the Dirac or Schr\"odinger equations of quantum mechanics  \cite{Mohr,Khan,Smith,Koukoutsis}.
Recent studies on a fully unitary representation of Maxwell equations in a simple dielectric have been the basis of qubit lattice algorithms (QLA) suitable for implementing on quantum computers \cite{Koukoutsis,Vahala1,Ram,unpublished}.

Current quantum computers are optimized to unitary operations. Such operators  naturally occur for closed quantum systems obeying the Schr\"odinger equation
\begin{equation}\label{1}
i\pdv{\ket{\bol{\psi}}}{t}=\hat{H}(\bol{r})\ket{\bol{\psi}}\quad\text{with}\quad\hat{H}=\hat{H}^\dagger.
\end{equation}
where $\ket{\bol{\psi}}$ is the wave function and $\hat{H}$ is the Hermitian generator of dynamics. Time evolution of $\ket{\bol{\psi}}$ is,
\begin{equation}\label{2}
\ket{\bol{\psi}(t)}=\hat{\mathcal{U}}\ket{\bol{\psi}(0)},
\end{equation}
where $\hat{\mathcal{U}}=\exp{-it\hat{H}}$ is the unitary evolution operator. For simplicity we assume  $\hat{H}$ to be time-independent.

Classical, linear and energy conserving systems admit a Schr\"odinger representation analogous to Eq.\eqref{1}, where unitary evolution corresponds to energy conservation. Then, it is possible to simulate the classical unitary dynamics through a series of unitary quantum gates in pursuit of higher computational efficiency compared to classical algorithms.

However, realistic systems exhibit dissipation (energy loss), prohibiting a straightforward application to quantum computers. For these classical systems the Schr\"odinger-like evolution equation is still valid but the generator of the dynamics is now non-Hermitian, $\hat{H}\neq\hat{H}^\dagger$, and possibly time-dependent, breaking the unitary evolution.

Quantum systems which include dissipation and decoherence due to interactions with an environment are
classified as open systems. In QIS, open systems can be studied using Kraus representation for the
evolution of the open system density matrix $\rho_S$ \cite{Nielsen},
\begin{equation}\label{3}
\rho_S(t)=\sum_{\mu}\hat{K}_{\mu}\rho_S(0)\hat{K}^\dagger_{\mu},\quad\sum_{\mu}\hat{K}_{\mu}^\dagger\hat{K}_{\mu}=I.
\end{equation}
For an open system, initially in the state $\rho_S(0)$ interacting with a stationary environment, with density matrix $\rho_E=\ket{e_0}\bra{e_0}$, the Kraus operators can be defined as,
\begin{equation}\label{4}
\hat{K}_{\mu}=\mel{\mu}{\hat{\mathcal{U}}_{S+E}}{e_0},
\end{equation}
where $\hat{\mathcal{U}}_{S+E}$ in Eq.\eqref{4} is the unitary evolution operator for the closed, composite system-environment, initially in the state $\rho_S\otimes\rho_E$ and $\ket{\mu}$ is an orthonormal basis for the state space of the environment. The representation in Eq.\eqref{3} is completely positive (CP), trace preserving (TP), and can always describe the time-dependent density matrix of an open system $\rho_S$ \cite{Tong}. 

In recent years, the concept of dilation  theory of operators has been used to embed the non-unitary evolution of open quantum systems into
the unitary framework appropriate for quantum computers \cite{Hu,Hubisz,Schlimgen,Schlimgen2,Shen}. In these studies,  a non-unitary
operator is considered as a projection of a unitary operator in an extended Hilbert space \cite{Moshe}. This extension of
the Hilbert space requires introducing ancillary qubits in order to appropriately represent the extended closed system.
Following this line of thought, the authors in Refs.\cite{Tip,Figotin,Cassier} have proven that Maxwell equations in a passive, dissipative and dispersive medium have a Hermitian Schr\"odinger structure, as in Eq.\eqref{1}, by extending the Hilbert space of the primary electromagnetic fields $\bol{E}, \bol{H}$ with appropriate set of auxiliary fields that are derived using functional analysis techniques. 
However, the resulting Hermitian Hamiltonian operator $\hat{H}$ has a complex structure that is not suitable for implementing on a quantum computer. Therefore, for a quantum implementation of Maxwell equations in lossy and dispersive media  a quantum representation that is compatible with the principles of QIS is required.

In this paper, we develop probabilistic dilation algorithms for simulating electromagnetic wave propagation in dissipative and dispersive electromagnetic media by expressing the non-unitary Suzuki-Trotter evolution of Maxwell equations within the framework of quantum channels.

In Sec.\ref{sec:2.1}, we formulate a quantum Schr\"odinger equation
representation of classical Maxwell equations in dispersive media by introducing auxiliary electromagnetic fields which are related to the wave polarization and to the polarization density current. In Sec.\ref{sec:2.2}, we examine the appearance of dissipation in the Maxwell-Schr\"odinger equation which leads to an anti-Hermitian component within the  Hamiltonian, generating  non-unitary Suzuki-Trotter evolution dynamics.

In Sec.\ref{sec:3} we formulate the classical electromagnetic dissipation as a post-selective augmented quantum amplitude damping-type channel \cite{Hubisz}. The respective set of Kraus operators form the basis of the dilation model with one ancillary qubit as the environmental state. In Secs.\ref{sec:3.1} and \ref{sec:3.2}, a quantum circuit for the probabilistic simulation of dissipative dynamics is presented along with the implementation scaling for the system-environment unitary operator into elementary quantum gates. Specifically, in Secs.\ref{sec:3.1.1}-\ref{sec:3.1.3} considerations about the  optimal time step and the probability of success for the post-selection are discussed, in addition to the post-selection complexity overhead for a total time-evolution with simulation error $\varepsilon$. Section \ref{sec:3.3} provides insights on an alternative deterministic, evaluation of the non-unitary dynamics based on the positive only non-linear quantum channels. We consider some expected outcomes from this approach. Finally, in Sec.\ref{sec:4}, we show the means to reduce the implementation cost of the dilated unitary evolution of dissipative systems using the Linear Combination of Unitaries (LCU) method that was firstly introduced in \cite{Long} and has been further developed  and integrated in contemporary conventional unitary quantum algorithms \cite{Childs} and non-unitary quantum algorithms for non-Hermitian or open quantum systems \cite{Zheng}. Relying on the LCU method, in Sec.\ref{sec:4.1} we further reduce the implementation cost for a simple dielectric medium with homogeneous dissipation rate and estimate the total circuit depth for the different evolution time-scales in the case of weak dissipation.

\section{Quantum representation of classical Maxwell equations in dispersive media}\label{sec:2}

We consider a six-vector formulation of the classical electromagnetic fields $\bol{u}=(\bol{E},\bol{H})^T$ and their
respective intensities $\bol{d}=(\bol{D},\bol{B})^T$ given by constitutive equations for a general dispersive medium \cite{Koukoutsis},
\begin{equation}\label{6}
\bol{d}(\bol{r},t)=\hat{W}(\bol{r})\bol{u}(\bol{r},t)+\int_0^t\hat{G}(\bol{r}, t-\tau)\bol{u}(\bol{r},\tau)d\,\tau.
\end{equation}
The instantaneous optical response of the medium is given by the $6\times6$ matrix
$\hat{W}=diag(\epsilon(\bol{r}),\mu(\bol{r}))$. The susceptibility kernel $\hat{G}$ yields
the linear dispersive response of the medium, accounting for both dissipation and memory effects.

The corresponding source-free classical Maxwell equations are,
\begin{align}\label{5}
   \div{\bol{d}(\bol{r}, t)}=0, \quad
   i\pdv{\bol{d}(\bol{r}, t)}{t}=\hat{M}\bol{u}(\bol{r},t),
\end{align}
where the Maxwell operator $\hat{M}$,
\begin{equation}\label{7}
\hat{M}=i\begin{bmatrix}
0&\curl\\
-\curl&0
\end{bmatrix}
\end{equation}
is Hermitian in $L^2(\mathcal{V}\subseteq\mathbb{R}^3,\mathbb{C})$ under the imposed Dirichlet boundary condition,
\begin{equation}\label{8}
\bol{n} (\bol{r})\times \bol E=0\,\,{\rm on\,\, the \,boundary}\,\,\partial\mathcal{V},
\end{equation}
where $\bol{n}$ is the outward pointing normal at the boundary.
The form of the constitutive relation \eqref{6} satisfies five physical postulates \cite{Roach}: determinism, linearity,
causality, locality in space, and invariance under time translations. In an inhomogeneous  and dispersion-less medium, the unitary form of Maxwell equations has been obtained in \cite{Koukoutsis}.

For a medium that is dispersive and dissipative and is described by a scalar permittivity and permeability,
the general constitutive relations in the frequency domain are \cite{Cassier},
\begin{align}
\epsilon(\bol{r},\omega)&=\epsilon_0\Big(1+\sum_{l=1}^{N_e}\frac{\Omega^2_{e,l}(\bol{r})}{\omega^2_{e,l}(\bol{r})-i\gamma_{e,l}(\bol{r})\omega-\omega^2}\Big),\label{9}\\
\mu(\bol{r},\omega)&=\mu_0\Big(1+\sum_{l=1}^{N_m}\frac{\Omega^2_{m,l}(\bol{r})}{\omega^2_{m,l}(\bol{r})-i\gamma_{m,l}(\bol{r})\omega-\omega^2}\Big).\label{10}
\end{align}
Equations \eqref{9} and \eqref{10} describe the phenomenological Lorentz oscillator response model of bound charges in materials. $\Omega_{e}$ and $\Omega_m$ are the characteristic frequencies of the oscillator, whereas $\omega_e$, $\omega_m$ are the resonant frequencies. The high frequency response of the medium is that of the vacuum with permittivity $\epsilon_0$ and permeability $\mu_0$. Finally, the $\gamma_{e,l}(\bol{r})$, $\gamma_{m,l}(\bol{r})\geq0$ are the relaxation-dissipation rates. If $\gamma_e=0$ and $\gamma_m=0$,  the medium is dispersive but lossless and for $\omega_e=0$ and $\omega_m=0$, we retrieve the Drude model for a simple metal.
The connection between the susceptibility kernel $\hat{G}$ in Eq.\eqref{6} and the constitutive relations \eqref{9},\eqref{10} in the frequency domain is,
\begin{equation}\label{11}
\hat{G}(\bol{r},t)=\frac{1}{2\pi}\int_{-\infty}^\infty\begin{bmatrix}
\epsilon({\bol{r},\omega})-\epsilon_0 & 0\\
0&\mu({\bol{r},\omega})-\mu_0
\end{bmatrix}e^{-i\omega t} \, d\omega
\end{equation}

Below, in Sec.\ref{sec:2.1}, we re-express the time dependent Maxwell equations \eqref{5} for
a Lorentz medium as a Schr\"odinger equation with Hermitian structure. When the medium is lossless, the unitary evolution, as expected, represents the conservation of the electromagnetic energy. The effect of dissipation is considered in Sec.\ref{sec:2.2}.

\subsection{Schr\"odinger representation of Maxwell equations in a lossless Lorentz medium}\label{sec:2.1}
We define the following auxiliary fields \cite{Cassier},
\begin{equation}\label{12}
\mathbb{P}_l(\bol{r}, t)=\frac{1}{2\pi}\int_0^t\int_{-\infty}^\infty\frac{e^{-i\omega(t-\tau)}}{\omega^2_{e,l}-\omega^2}\bol{E}(\bol{r},\tau)d\,\omega d\,\tau,
\end{equation}
and,
\begin{equation}\label{13}
\mathbb{M}_l(\bol{r}, t)=\frac{1}{2\pi}\int_0^t\int_{-\infty}^\infty\frac{e^{-i\omega(t-\tau)}}{\omega^2_{m,l}-\omega^2}\bol{H}(\bol{r},\tau)d\,\omega d\,\tau
\end{equation}
which are related to the electric $\bol{P}$ and magnetic $\bol{M}$ polarizability of the medium,
\begin{equation}\label{14}
\bol{P}(\bol{r}, t)=\epsilon_0\sum_{l=1}^{N_e}\Omega^2_{e,l}\mathbb{P}_l,\quad \bol{M}(\bol{r}, t)=\mu_0\sum_{l=1}^{N_m}\Omega^2_{m,l}\mathbb{M}_l.
\end{equation}
Maxwell equations \eqref{5}, together with the evolution equations for the auxiliary fields \eqref{12},\eqref{13} yield the following closed system of partial differential equations,
\begin{equation}\label{15}
\begin{aligned}
i\pdv{\bol{u}}{t} &= \hat{W}_0^{-1}\hat{M}\bol{u}-i\sum_{l=1}^{N}\hat{\Omega}_l^2\bol{\mathcal{P}}_{l,t},\\
i\pdv{\bol{\mathcal{P}}_l}{t} &=i\bol{\mathcal{P}}_{l,t}, \quad \quad l=1,2...N,\\
i\pdv{\bol{\mathcal{P}}_{l,t}}{t} &= i\bol{u}-i\hat{\omega}^2_l\bol{\mathcal{P}}_l, \quad \quad l=1,2...N,
\end{aligned}
\end{equation}
where $\bol u=(\bol E, \bol H)^T$, $\bol{\mathcal{P}}_l=(\mathbb{P}_l,\mathbb{M}_l)^T$, $N=max\{N_e,N_m\}$
and the diagonal matrices $\hat{\Omega}_l^2$ and $\hat{\omega}^2_l$ are,
\begin{equation}\label{16}
\hat{\Omega}_l^2=\begin{bmatrix}
\Omega^2_{e,l}&0\\
0& \Omega^2_{m,l}
\end{bmatrix},\quad\hat{\omega}^2_l=\begin{bmatrix}
\omega^2_{e,l}&0\\
0&\omega^2_{m,l}
\end{bmatrix}.
\end{equation}\\
Upon applying the Dyson transform \cite{Koukoutsis},
 \begin{equation}
\hat{\rho}=diag(\hat{W}_0^{1/2},\hat{W}_0^{1/2}\hat{\Omega}_l\hat{\omega}_l,W_0^{1/2}\hat{\Omega}_l),
 \end{equation}
to Eq.\eqref{15}, we obtain a Hermitian Schr\"odinger representation of Maxwell equations,
\begin{equation}\label{18}
\begin{split}
&i\pdv{}{t}\begin{bmatrix}
\hat{W}_0^{1/2}\bol{u}\\
\hat{W}_0^{1/2}\hat{\Omega}_l\hat{\omega}_l\bol{\mathcal{P}}_l\\
W_0^{1/2}\hat{\Omega}_l\bol{\mathcal{P}}_{l,t}
\end{bmatrix}=\begin{bmatrix}
c\hat{M} & 0 &-i\hat{\Omega}_l\\
0 &0&i\hat{\omega}_l\\
i\hat{\Omega}_l&-i\hat{\omega}_l&0
\end{bmatrix}\begin{bmatrix}
W_0^{1/2}\bol{u}\\
\hat{W}_0^{1/2}\hat{\Omega}_l\hat{\omega}_l\bol{\mathcal{P}}_l\\
W_0^{1/2}\hat{\Omega}_l\bol{\mathcal{P}}_{l,t}
\end{bmatrix},\\
& l=1,2,...,N,
\end{split}
\end{equation}
where $c=(\epsilon_0\mu_0)^{-1/2}$ is the speed of light in vacuum.
This evolution equation can be compactly written as,
 \begin{equation}\label{18a}
i\pdv{\bol{\psi}}{t}=\hat D\bol{\psi},\quad\hat D=\hat D^\dagger,
\end{equation}
with the initial condition $\bol\psi_0=(\hat{W}_0^{1/2}\bol u_0,0,0)^T$ and $\bol u_0=(\bol E_0, \bol H_0)^T$ is the initial electromagnetic field. Equation \eqref{18a} is called the Maxwell-Schr\"odinger equation due to its quantal form similar in mathematical properties to Schrodinger equation \eqref{1}. The state vector $\bol\psi$ includes all the physically relevant electromagnetic fields that are necessary to understand propagation and scattering of waves in a lossless Lorentz medium. A generalization of Eq.\eqref{18} to a tensorial Lorentz medium  can be found in \cite{Silveirinha}.

 A quantum implementation of unitary evolution operator $\exp{-it\hat D}$ that results from Eq.\eqref{18a}, has been delineated in \cite{Koukoutsis2} for the special case of magnetized plasma based on a qubit lattice algorithm (QLA). A QLA consists of an interleaved sequence of non-commuting collision $\hat{C}$ and streaming $\hat{S}$ operators that recover the Maxwell-Schr\"odinger equation \eqref{18} to second order diffusion scheme $\delta t\sim\delta^2$, $\delta\bol r=(\delta x,\delta y,\delta z)\sim\delta$. We define $\delta$ to be the Cartesian grid spacing between the $(N_x, N_y, N_z)$ number of grid points along the principal axes that cover the entire domain $\mathcal{V}$.

We express the  state vector $\bol\psi$, as a $n$-qubit superposition state,
where $n=\log_2{d}$ and $d =(6 +12N) N_x N_y N_z $ is the dimensionality of the state after discretization. The total number of qubits $n$ can be divided into two registers of $n_q=\log_2(6 +12N)$ and $n_p=\log_2( N_x N_y N_z)$ qubits, representing  the amplitude and spatial gridding, respectively. The advantage of QLA implementation is that  the collision operators $\hat C$, for the homogeneous case, act only locally as controlled rotations on the amplitude $n_q$ register; hence their implementation cost is $\textit{O}(n_q^2)=constant$, whereas the streaming $\hat S$ operators can be considered as a quantum walk in the $n_p$ register that can be implemented in $\textit{O}(n_p^2)$ single qubit and contolled-NOT (CNOT) operations \cite{Koukoutsis,Gourdeau}.

The unitary evolution of Eq.\eqref{18a} leads to the conservation of the extended electromagnetic energy $E_{total}$ which is the norm of the state vector $\norm{\bol\psi}^2$,
\begin{equation}\label{19}
\begin{split}
E_{total}(t)&=\frac{1}{2}\norm{\bol\psi}^2=\frac{1}{2}\hat{W}_0\int_\Omega\norm{\bol u}^2 d\,\bol r\\
&+\frac{1}{2}\hat{W}_0\sum_{l=1}^N\int_\Omega\hat{\Omega}_l^2\left(\hat{\omega}_l^2\norm{\bol{\mathcal{P}}_l}^2+\norm{\bol{\mathcal{P}}_{l,t}}^2\right) d\,\bol r.
\end{split}
\end{equation}
The first term on the right hand side of \eqref{19} is the vacuum electromagnetic energy,
\begin{equation}\label{20}
E_{el}(t)=\frac{1}{2}\int_\Omega\epsilon_0\left(\norm{\bol E}^2+\mu_0\norm{\bol H}^2\right) d\,\bol r\leq E_{total}(0)=E_{el}(0).
\end{equation}
Energy expression \eqref{19} is valid beyond the plane-wave, harmonic and semi-harmonic approximations for the 
fields as imposed by Landau and Brillouin \cite{Landau}.

\subsection{Dissipative Medium}\label{sec:2.2}
For a dissipative medium, in Eqs.\eqref{9} and \eqref{10}, $\gamma_{e,l}\geq0$ and $\gamma_{m,l}\geq0$. Consequently, the denominators in Eqs.\eqref{12} and \eqref{13} include the appropriate factors of $-i\gamma_{e,l}$ and $-i\gamma_{m,l}$, respectively. As a result, the Hermitian structure of the Maxwell-Schr\"odinger evolution equation \eqref{18a} is not preserved. While the first two Maxwell equations in \eqref{15} are not affected by dissipation,
the third equation now reads,
\begin{equation}\label{21}
i\pdv{\bol{\mathcal{P}}_{l,t}}{t}=i\bol{u}-i\hat{\omega}^2_l\bol{\mathcal{P}}_l - 
i\hat{\gamma}_l\bol{\mathcal{P}}_{l,t}, \quad \quad l=1,2...N,
\end{equation}
where,
\begin{equation}\label{22}
\hat{\gamma}_l=\begin{bmatrix}
 \gamma_{e,l} & 0\\
0 &  \gamma_{m,l}
\end{bmatrix}.
\end{equation}
The dissipative counterpart to Eq.\eqref{18a} has the form,
\begin{equation}\label{23}
i\pdv{\bol\psi}{t}=[\hat D-i\hat{D}_{diss}]\bol\psi.
\end{equation}
The diagonal matrix $\hat{D}_{diss}=diag(0,0,\hat{\gamma}_l)$ is Hermitian and positive definite ($\gamma_{e,l} \geq 0$ and $\gamma_{m,l} \geq 0)$ so the anti-Hermitian term $-i\hat{D}_{diss}$ is purely dissipative. As a consequence, the non-Hermitian generator of dynamics $\hat D-i\hat{D}_{diss}$ in Eq.\eqref{23} possesses both real eigenvalues as well as complex eigenvalues but with negative imaginary part, indicating the absence of any global Parity-Time ($\mathcal{PT}$) \cite{Bender} or pseudo-Hermitian structure \cite{Mostafazadeh}. An example of $\mathcal{PT}$-symmetry in electrodynamics is two coupled optical systems with balanced gain and loss \cite{Zyablovsky} whereas a paradigm on unbroken pseudo-Hermitian structure of Maxwell equations in passive media for wave propagation can be found in \cite{Koukoutsis}. Moreover, we would like to address that even if $\mathcal{PT}$ symmetry was present, by permitting gain (now some of $\gamma_{e,l}$ and $\gamma_{m,l}$ can be negative) in constitutive relations Eqs.\eqref{9} and \eqref{10}, an unbroken $\mathcal{PT}$ region where the non-Hermitian operator $\hat D-i\hat{D}_{diss}$ possesses only real eigenvalues will be present at discrete frequencies \cite{Zyablovsky}, due to Kramers-Kroning causality relations \cite{Landau}.

Thus, the resulting evolution operator $\hat{\mathcal{U}}(t)=\exp{-it[\hat D-i\hat{D}_{diss}]}$ is strictly non-unitary. On the face of it, this non-unitarity poses a challenge for implementation on a quantum computer, a challenge we will tackle from this point onward.

For an infinitesimal time step $\delta t$, starting at $t=0$, a first order Suzuki-Trotter approximation
of the non-unitary operator $\hat{\mathcal{U}}(t)$ yields,
\begin{equation}\label{24}
\exp{-i\delta t[\hat D-i\hat{D}_{diss}]}=e^{-i\delta t\hat D}e^{-\delta t\hat{D}_{diss}}+ \textit{O}(\delta t^2).
\end{equation}
This allows us to separate out the non-unitary term $\exp{-\delta t\hat{D}_{diss}}$. 

The diagonal dissipative operator $\hat{D}_{diss}$ contains at most $6N$ positive elements $\left( \gamma_{e,l},\,\gamma_{m,l} \right)$,
$l = 1, \dots N$. Then,
\begin{equation}\label{25}
\exp{-\delta t\hat{D}_{diss}}=\hat{K}_0=diag(I_{6\times6}, I_{6N\times6N}, \hat{\Gamma}),
\end{equation}
where,
\begin{equation}\label{26}
\hat{\Gamma}=\begin{bmatrix}
e^{-\delta t\gamma_{e,l}}I_{3\times3} &0\\
0& e^{-\delta t\gamma_{m,l}}I_{3\times3}
\end{bmatrix}.
\end{equation}
The dimensions of the diagonal sub-matrix $\hat{\Gamma}$ are $6N\times6N$. The non-unitary operator $\exp{-\delta t\hat{D}_{diss}}$ has been denoted as $K_0$ because it will represent the first Kraus operator in the quantum channel description Eq.\eqref{3} of classical dissipation that follows in the next section.

\section{Dissipation in the context of quantum channels}\label{sec:3}

In the density matrix framework, the Suzuki-Trotter evolution
\eqref{24} is,
\begin{equation}\label{28}
\Bar{\rho}(\delta t)=e^{-i\delta t\hat D}\hat{K}_0\rho(0)\hat{K}_0^\dagger e^{i\delta t\hat D},
\end{equation}
where the initial density matrix $\rho(0)$ is,
\begin{equation}\label{27}
\rho(0)=\ket{\bol\psi_0}\bra{\bol\psi_0},\quad \ket{\bol\psi_0}=\frac{1}{\sqrt{E_0}}\sum_{j=0}^{d-1}\psi_{0{j}}\ket{j}.
\end{equation} 
From Eq.\eqref{20}, the initial energy is $E_0=\braket{\bol{\psi}_0}=\frac{1}{2}\sum_{j}\epsilon_0E^2_{0j}+\mu_0H^2_{0j}$. By construction, the initial state $\ket{\bol\psi_0}$ is a pure state. It should be noted that $\Bar{\rho}(\delta t)$ is not a proper quantum mechanical density matrix, as the 
operation $\hat{K}_0\rho(0)\hat{K}_0^\dagger$ is not trace preserving (TP).
Non-TP quantum channels emerge when a measurement is performed in the environment and selecting over a specific outcome \cite{Nielsen}. Consequently, we can think of classical dissipation as a post-selective outcome from the interaction between a quantum represented lossless system and an unspecified environment. Accordingly, to retrieve the TP property we augment Eq.\eqref{28} with the term, 
\begin{equation}\label{29}
\hat{K}_1\rho(0)\hat{K}_1,
\end{equation}
where the second Kraus operator $\hat{K}_1$ satisfies $\hat{K}^\dagger_1\hat{K}_1=I_{d\times d}-\hat{K}_0^\dagger\hat{K}_0$, and has an upper-triangular form,
\begin{equation}\label{30}
\hat{K}_1=\begin{bmatrix}
0&\sqrt{I_{r\times r}-\hat{\Gamma}^2}\\
0&0
\end{bmatrix},
\end{equation}
with $r=6NN_xN_yN_z$ being the dimension of the dynamic space occupied by dissipation.
The operator $\hat{K}_1$ corresponds to a transition -- a quantum jump from the dissipative state of interest to a different state. The  operators $\hat{K}_0$ and $\hat{K}_1$ are the multi-dimensional analogs of the amplitude damping 
channel operators \cite{Nielsen}. The augmented quantum dissipative evolution for the open quantum system is,
\begin{equation}\label{31}
\rho_{aug}(\delta t)=e^{-i\delta t\hat D}\rho_{diss}(\delta t)e^{i\delta t\hat D} ,
\end{equation}
where,
\begin{equation}\label{32}
\rho_{diss}(\delta t)=\hat{K}_0\rho(0)\hat{K}_0^\dagger+\hat{K}_1\rho(0)\hat{K}_1^\dagger.
\end{equation}
The operator $e^{-i\delta t\hat D}$ has been defined in Eq.\eqref{18}.

The constructed linear CPTP quantum channel in Eqs.\eqref{31},\eqref{32} that supports dissipation describes the evolution of the linear dynamics of an open quantum system, generated by the effective  Hamiltonian $\hat H_{eff}$,
\begin{equation}\label{extra1}
\hat H_{eff}=\hat D-i\hat{L}^\dagger\hat L,\quad\hat L=\begin{bmatrix}
0 & \sqrt{\hat{\gamma}_r}\\
0 &0
\end{bmatrix}.
\end{equation}
The diagonal matrix $\hat{\gamma}_r$, defined in Eq.\eqref{22}, represents dissipation in the $r$-dimensional subspace. The operator 
$\hat L$ is called the Lindblad jump operator \cite{Lindblad}. The generated Gorini–Kossakowski–Sudarshan–Lindblad (GKSL) master equation \cite{Lindblad,Gorini} is then,
\begin{equation}\label{extra2}
\pdv{\rho}{t}=-i\hat H_{eff}\rho+i\rho\hat H_{eff}^\dagger+2\hat L\rho\hat L^\dagger.
\end{equation}
For an infinitesimal time evolution $0\to\delta t$, the density matrix evolution, to first order in $\delta t$, can be generated through the master equation \eqref{extra2} for the classical, non-Hermitian operator,
\begin{equation}\label{extra3}
\rho(\delta t)=\hat{E}_0\rho(0)\hat{E}_0^\dagger+\hat E_1\rho(0)\hat E_1^\dagger,
\end{equation}
with
\begin{equation}\label{extra4}
\hat{E}_0=I_{d\times d}-i\delta t\hat D-\delta t\hat D_{diss},\quad \hat E_1=\sqrt{2\delta t}\hat L.
\end{equation}
By expanding in a Taylor series the Kraus operators in the augmented evolution Eq.\eqref{31} we obtain,
\begin{equation}\label{extra5}
\rho_{aug}(\delta t)=\rho(\delta t)+\textit{O}(\delta t^2).
\end{equation}
Equation \eqref{extra5} confirms that treating the classical non-Hermitian operator $\hat D-i\hat{D}_{diss}$ as a quantum effective Hamiltonian generates, to first order, the same dynamics with the quantum channel of Eqs.\eqref{31},\eqref{32}. In the following section, we establish that this minimal augmented form is sufficient to capture the classical dissipative dynamics.

\subsection{The algorithm}\label{sec:3.1}
Since the set of Kraus operators $\hat{K}_0$ and $\hat{K}_1$ in Eq.\eqref{32} define a linear CPTP quantum channel, they are contractions \cite{Hu}. Thus, a guaranteed minimal unitary dilation $\hat{\mathcal{U}}_{diss}$ can be formulated for the Suzuki-Trotter evolution \eqref{24} of the open quantum system,
\begin{equation}\label{33}
\hat{\mathcal{U}}_{diss}=\begin{bmatrix}
\hat{K}_0&-\hat{K}_1^\dagger\\
\hat{K}_1&\hat{\mathcal{X}}\hat{K}_0\hat{\mathcal{X}}
\end{bmatrix},
\end{equation}
without resorting to block-encoding techniques that introduce a failure error. The operator $\hat{\mathcal{X}}$ is an appropriate extension of the Pauli $\hat{X}$ operator to $d$-dimensions. The unitary $\hat{\mathcal{U}}_{diss}$ is a $2d\times2d$ matrix operator acting on $n+1$ qubits. The ancillary qubit represents the environment; the lossless system together with the environment form a closed conservative system that evolves under the unitary operator $\hat{\mathcal{U}}_{diss}$. This minimal dilation is related to the Sz. Nagy dilation \cite{Moshe} of $\hat{K}_0$ 
by a rotational transformation. 

\begin{figure}[ht]
    \centering
   \includegraphics{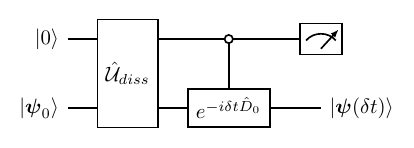}
   \caption{Quantum circuit for simulation of the non-unitary classical evolution \eqref{24} in a dissipative and dispersive electromagnetic medium.}
   \label{fig:1}
\end{figure}

We consider the qubit environment to be stationary with initial density matrix $\rho_E=\ket{0}\bra{0}$, and the initial state in the dilated space 
$\ket{\bol{\Psi}_0}$ to be separable,
\begin{equation}\label{34}
\ket{\bol{\Psi}_0}=\ket{0}\ket{\bol\psi_0}.
\end{equation}
The action of $\hat{\mathcal{U}}_{diss}$ on the composite initial state \eqref{34} yields,
\begin{equation}\label{35}
\ket{0}\hat{K}_0\ket{\bol\psi_0}+\ket{1}\hat{K}_1\ket{\bol\psi_0}.
\end{equation}
Next, we apply a controlled $e^{-i\delta t\hat D}$ operation to state \eqref{35} with respect to the $0$-bit environment qubit, leading to the composite state,
\begin{equation}\label{36}
\ket{0}e^{-i\delta t\hat D}\hat{K}_0\ket{\bol\psi_0}+\ket{1}\hat{K}_1\ket{\bol\psi_0}.
\end{equation}
Subsequently, a projective measurement in the first qubit with operator $P_0=\ket{0}\bra{0}\otimes I_{d\times d}$ followed by tracing out the environment, leads to the non-unitary Suzuki-Trotter evolution equation \eqref{24} for the lossy, dispersive medium, up to a normalization factor. The steps in Eqs.\eqref{34}-\eqref{36} along with the post-selection of the output state are illustrated in the quantum circuit in Fig.\ref{fig:1}.

\subsubsection{State preparation}\label{sec:3.1.1}
Preparation of the initial state $\ket{\bol\Psi_0}$ of Eq.\eqref{34} in the dilated space requires the preparation of the initial electromagnetic state $\ket{\bol\psi_0}$,
\begin{equation}\label{preparation}
\ket{0}^{\otimes n}\to\ket{\bol\psi_0}=\ket{\bol u_0}=\ket{\bol E_0,\bol B_0}=\sum_{k=0}^{r/N-1}\psi_{0,k}\ket{k}.
\end{equation}
Due to causality, the initial state \eqref{preparation} is sparse with $r/N=\textit{O}(2^{n-1}/N)$ non-zero elements in general. As a result, based on \cite{Zhang}, the initial state \eqref{preparation}  can be implemented with a circuit depth of $\Theta[n-1+\log_2{(\frac{n}{N})}]$ using $\textit{O}(2^{n-1}n^2/N)$ ancillary qubits. For studying the propagation of an initial state that is localized in space, for example, Gaussian pulse or wave-packet, the requirements are further reduced as we need fewer qubits to describe the initial state in the discretized space $\mathcal{V}$.

\subsubsection{Post-selection complexity}\label{sec:3.1.2}
To first order in $\delta t$, the probability $p_0(\delta t)=\mel{\bol\psi_0}{\hat{K}^2_0}{\bol\psi_0}$ for a successful post-selection is,
\begin{equation}\label{new1}
 p_0(\delta t)=1-2\delta t\sum_{q=d-r}^{d-1}\gamma_q\frac{\abs{\psi_{0q}}^2}{E_0}.
\end{equation}
where $p_0 (\delta t)$ is bounded $p_{0min} (\delta t) \le p_0 (\delta t) \le p_{0max} (\delta t)$,
\begin{align}
p_{0min}(\delta t)&=1-2\gamma_{max}\delta t\sum_{q=d-r}^{d-1}\frac{\abs{\psi_{0q}}^2}{E_0}\label{new2},\\
p_{0max}(\delta t)&=1-2\gamma_{min}\delta t\sum_{q=d-r}^{d-1}\frac{\abs{\psi_{0q}}^2}{E_0}\label{new3},
\end{align}
with $r=6NN_xN_yN_z$, $\gamma_{max}={\rm max}\{\gamma_{e,l},\gamma_{m,l} \}$, and
$\gamma_{min}= {\rm min} \{\gamma_{e,l},\gamma_{m,l} \}$, $l = 1, \dots, r$.
From \eqref{new1}, the optimal time-step $\delta t$ for a high success post-selection out of the output state \eqref{36} requires,
\begin{equation}\label{new4}
\delta t<<\frac{1}{2\sum_{q=d-r}^{d-1}\gamma_q\frac{\abs{\psi_{0q}}^2}{E_0}}.
\end{equation}
Based on \eqref{new2}, the upper bound on $\delta t$ is,
\begin{equation}\label{new5}
\Delta t_{diss}=\frac{1}{2\gamma_{max}\sum_{q=d-r}^{d-1}\frac{\abs{\psi_{0q}}^2}{E_0}},
\end{equation}
with $\delta t<<\Delta t_{diss}$, where the time step $\Delta t_{diss}$ corresponds to the fast time scale $1/\gamma_{max}$ associated with dissipation -- 
since $\sum_{q=d-r}^{d-1}\frac{\abs{\psi_{0q}}^2}{E_0} \sim r/d \sim 1/2$,  $\Delta t_{diss}\sim 1/\gamma_{max}$. 
Thus, for an accurate modeling of dissipation we have to select a simulation time step that is smaller than the fastest dissipative time scale. We have realized this physical conclusion solely from the quantum operational requirement of a highly successive post-selection process.

The non-normality of operator $\hat D-i\hat{D}_{diss}$ dictates that besides the dissipation time scale $\Delta t_{diss}$ which is dominant for large simulation time, there is also a time scale $\Delta t_{uni}=1/\lambda_{max}$ associated with the Hermitian part $\hat{D}$ with $\lambda_{max}=\lambda_{max}(n,\Omega_{e,m},\omega_{e,m})$ the largest eigenvalue of operator $\hat{D}$ \cite{Okuma}, depending on the number of qubits $n$ and on the parameters present in operator $\hat{D}$. This time scale is dominant for small simulation time.

The quantum circuit in Fig.\ref{fig:1} can be interpreted as a building block for a quantum simulation of total time $t_{total}=N_t\delta t$ in two different ways \cite{Shen}. The first one utilizes the extra qubit globally and applies the quantum circuit of Fig.\ref{1}, $N_t$ consecutive times. Then, a single post-selection on the total output state is enough to obtain the desired normalized non-unitary evolution,
\begin{equation}\label{evolution state}
\ket{\bol{\psi}(t_{total})}=\frac{e^{-it_{total}(\hat{D}-i\hat{D}_{diss})}\ket{\bol\psi_0}}{\norm{e^{-it_{total}(\hat{D}-i\hat{D}_{diss})}\ket{\bol\psi_0}}}.
\end{equation}
However, taking into consideration the probability $p_{total}$ for a successful post-selection at the final stage, this decays exponentially \cite{Okuma} as
\begin{equation}\label{decay probabilityt}
p_{total}=\norm{e^{-it_{total}(\hat{D}-i\hat{D}_{diss})}\ket{\bol\psi_0}}^2.
\end{equation}
Thus, Eq.\eqref{decay probabilityt} implies that we need an exponentially large number, $N_t/p_{total}$, of repetitions of the quantum circuit in Fig.\ref{fig:1}, for obtaining the state \eqref{evolution state} to within a desired error. This picture refers to monitoring only the continuous evolution of a quantum trajectory by disposing the trajectories with quantum
jumps with a global post-selection process \cite{Okuma,Daley}. The second implementation path for the total evolution is a repetition of Fig.\ref{fig:1} quantum circuit to be followed by a a post-selection with success probability $p_0$ each time. This local re-usage of the extra qubit in each step exploits the magnitude optimization Eq.\eqref{new5} of probability $p_0$ relying on the dissipation characteristics. The post-selection complexity for the total evolution now reads $N_t/p^{N_t}_0$.

While the global method incorporates only one post-selection the exponentially small success rate of it, Eq.\eqref{decay probabilityt}, renders it extremely costly in terms of resources for large time-scale simulation. Conversely, the local method demands of $N_t$ post-selections which introduces a readout error, but the overall success rate is realizable after suitable selection of the $\delta t$ time step.

\subsubsection{Time complexity}\label{sec:3.1.3}
As described in Sec.\ref{sec:3.1.2}, for the local simulation method with the intermediate post-selections to be efficiently applied, a suitable selection of $\delta t$ has to made in relation with the physical time scales $\Delta t_{diss}$ and $\Delta t_{uni}$ to capture the transient physical phenomena \cite{Okuma}. However, a fine-meshing in the temporal domain directly affects the number of Trotter steps $N_t$ required for a complete simulation time $t_{total}=N_t\delta t$ within an error $\varepsilon$. After $N_t$ Trotter repetitions the error of the resulting (without considering the post-selection complexity overhead) non-unitary evolution $(e^{-i\delta t\hat D}e^{-i\delta t\hat D_{diss})^{N_t}}$  is \cite{Childs2},
\begin{equation}\label{new6}
\varepsilon\sim\frac{\gamma_{max}\lambda_{max}t^2_{total}}{N_t}\exp{\frac{2(\lambda_{max}+\gamma_{max})}{N_t}t_{total}}
\end{equation}
reflecting the trade-off between the smallness of the time step $\delta t$ and the number of repetitions $N_t$ for the desired error scaling.

Considering a more sophisticated and higher order product formulas \cite{Childs2} can lead to an optimized selection between the quantities $\varepsilon, N_t,\delta t$ and $p_0$.

\subsection{Implementation of the \texorpdfstring{$\hat{\mathcal{U}}_{diss}$}{TEXT} operator}\label{sec:3.2}
The explicit form of $\hat{\mathcal{U}}_{diss}$ in Eq.\eqref{33} is,
\begin{equation}\label{37}
\hat{\mathcal{U}}_{diss}=\begin{bmatrix}
I_{(d-r)\times(d-r)} &0 & 0 &0\\
0 &\hat{\Gamma} & -\sqrt{I_{r\times r}-\hat{\Gamma}^2} &0\\
0 & \sqrt{I_{r\times r}-\hat{\Gamma}^2} & \hat{\Gamma} &0\\
0 &0&0& I_{(d-r)\times(d-r)}
\end{bmatrix},
\end{equation}
where the diagonal operator $\hat{\Gamma}$ is given in Eq.\eqref{26}. 
Setting $\cos{\theta_l/2}=\hat{\Gamma}_{ll}$, we can decompose $\hat{\mathcal{U}}_{diss}$ into $r$ two-level unitary $y$-rotations, $\hat{\mathcal{R}}_y(\theta_l)$, 
acting on $n+1$ qubits,
\begin{equation}\label{38}
\hat{\mathcal{U}}_{diss}=\prod_{l=1}^r\hat{\mathcal{R}}_y(\theta_l).
\end{equation}
Hence, to leading order, we can implement $\hat{\mathcal{U}}_{diss}$ in $\textit{O}(rn^2)$ CNOTs and a single qubit rotations $\hat{R}_y(\theta_l)$. Since $d=(6+12N)N_xN_yN_z=6N_xN_yN_z+2r=2^n$ then $r=2^{n-1}(1-\frac{1}{1+2N})$, and the implementation of $\hat{\mathcal{U}}_{diss}$ is achieved using
$\textit{O}(2^{n-1}n^2)$ simple gates.

The decomposition of $\hat{\mathcal{U}}_{diss}$ into a product of two-level rotations Eq.\eqref{38} acting on $n+1$ qubits is related to multiplex rotations. Following \cite{Mottonen}, a multiplex rotation acting on $n+1$ qubits can be implemented  in $\textit{O}(4^{n+1})$ CNOTs and
single qubit $\hat{R}_y$ rotations. The two crucial aspects that enable us to implement the product structure Eq.\eqref{38} in an exponentially better way are, firstly that the dissipation subspace $\mathcal{H}_r$ has a smaller dimension $r<d$ compared to the overall system, and secondly the Hilbert space $\mathcal{H}_r$ structure of the classical system enables decomposition into separate spaces through a direct sum, $\mathcal{H}_r=\bigoplus_{l=1}^r\mathcal{H}_l$. The latter is not 
always possible in quantum systems and is the cause of the Strinespring scaling \cite{Stinespring} in the amplitude damping dilation of \cite{Hubisz} 
compared to our one-qubit amplitude damping dilation.

\subsection{Connection with non-linear quantum channels}\label{sec:3.3}

The quantum circuit of Fig.\ref{fig:1} produces the normalized classical evolved density matrix, 
\begin{equation}\label{extra9}
\rho(\delta t)=\frac{e^{-i\delta t\hat D}\hat K_0\rho(0)\hat{K}_0^\dagger e^{i\delta t\hat D}}{\mel{\bol\psi_0}{\hat{K}^2_0}{\bol\psi_0}},
\end{equation}
after a deterministic evolution followed by a successful post-selective projective measurement that introduces a probabilistic aspect to our algorithm along with the pertinent errors (see the discussion at the end of Sec.\ref{sec:3.2}).

It is possible to obtain the desired result of Eq.\eqref{extra9} purely deterministically by employing a non-linear in normalization only (NINO) evolution quantum channel\cite{Kowalski,Rembielinski,Geller1,Geller2},
\begin{equation}\label{extra10}
\rho_0\to\rho(\delta t)=\frac{\Phi(\rho_0)}{Tr[\Phi(\rho_0)]},
\end{equation}
where the linear map $\Phi$ in Eq.\eqref{extra10} is,
\begin{equation}
\Phi(\rho_0)=e^{-i\delta t\hat D}\hat K_0\rho(0)\hat{K}_0^\dagger e^{i\delta t\hat D}.
\end{equation}
NINO channels are weakly non-linear quantum channels that have found applications in quantum information processes such as fast quantum purification \cite{Kowalski}, amplification \cite{Geller2}, mixed state evolution in $\mathcal{PT}$ symmetric systems \cite{Brody} and state discrimination \cite{Geller1} as well as implicit application in the theory of classical dynamical systems \cite{Das}.

The generated non-linear extension of GKSL equation for the dynamics of the open quantum system is,
\begin{equation}\label{extra6}
\pdv{\rho}{t}=-i\hat{H}_{eff}\rho=i\rho\hat H_{eff}-\rho\hat D_{diss}+2R\rho,
\end{equation}
with
\begin{equation}\label{extra7}
R=Tr(\hat{D}_{diss}\rho),
\end{equation}
where $Tr$ denotes the trace operation. From classical perspective, the quantity $R$ is the instantaneous Lyapunov exponent \cite{Das} that translates into relaxation exponent in our case due to the existence of dissipation.

Comparing the master equation Eq.\eqref{extra2} and its non-linear counterpart Eq.\eqref{extra7} we deduce two important differences: $(1)$ Equation \eqref{extra7} conserves the trace non-linearly without any jump operators, in contrast with Eq.\eqref{32} where the jump operator $\hat{L}$ has been introduced in Eq.\eqref{extra1}. Thus, Eq.\eqref{extra9} is derived deterministically without the post-selection error overhead. $(2)$ The linear CPTP channel preserves the trace though the condition $\hat{D}_{diss}=\hat{L}^\dagger\hat{L}$ which implies that the non-Hermitian operator $-i\hat{D}_{diss}$ has to describe only dissipative processes (all eigenvalues of $\hat{D}_{diss}$ are positive). NINO channels support both dissipation and amplification, by relaxing the (CP) condition to simple positivity (P) \cite{Geller1}. Hence, it may be possible to simulate active electromagnetic media and meta-materials with combined gain and loss exploiting the PTP NINO channels.

\section{An optimized approach}\label{sec:4}

In the previous section, we succeeded in showing the pathway to convert the non-unitary, diagonal, dissipative Kraus operator 
$\hat{K}_0=\exp{-\delta t\hat{D}_{diss}}$ into $\textit{O}(2^{n-1}n^2)$ elementary unitary gates, based on the constructed 
interconnection between dissipative and post-selective open quantum systems and classical dissipation.
A different approach is to refrain from associating classical dissipation with a quantum process but treat $\hat{K}_0$ with the LCU method \cite{Long,Childs,Zheng}. Specifically, 
$\hat{K}_0$ can be written as a sum of two unitary matrices,
\begin{equation}\label{46}
\hat{K}_0=\frac{1}{2}(\hat{K}_{0z}+\hat{K}_{0z}^\dagger),
\end{equation}
where,
\begin{equation}\label{47}
\hat{K}_{0z}=\begin{bmatrix}
I_{(d-r)\times(d-r)}&0\\
0& e^{-i\theta_l/2}
\end{bmatrix},\quad l=1,2,...,r.
\end{equation}
As in Sec.\ref{sec:3.2}, we have set $\cos{\theta_l/2}=\hat{\Gamma}_{ll}$. It is important to realize that the unitary 
components in Eq.\eqref{46} remain diagonal. In order to apply the LCU method (Lemma 6 in \cite{Childs}), we need one auxiliary qubit as in the dilation process of Sec.\ref{sec:3}. We introduce the unitary operators,
\begin{align}
\hat{U}_{prep}&:\ket{0}\to\frac{1}{\sqrt{2}}(\ket{0}+\ket{1})\label{48},\\
\hat{U}_{select}&=\ket{0}\bra{0}\otimes\hat{K}_{0z}+\ket{1}\bra{1}\otimes\hat{K}_{0z}^\dagger\label{49},
\end{align}
where $\hat{U}_{prep}=\hat{H}$ is the Hadamard gate. We can probabilistically implement $\hat{K}_0$ using the unitary dilated operator,
\begin{equation}\label{50}
\hat{\mathcal{U}}_{diss}=(\hat{H}\otimes I_{d\times d})\hat{U}_{select}(\hat{H}\otimes I_{d\times d}).
\end{equation}

\begin{figure}[ht]
   \centering
  \includegraphics[scale=0.95]{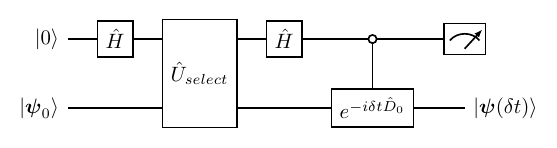}
   \caption{Quantum circuit for simulation of the non-unitary classical evolution \eqref{24} in a dissipative and dispersive medium using the LCU method.}
  \label{fig:2}
\end{figure}

The action of $\hat{\mathcal{U}}_{diss}$ on the initial state $\ket{0}\ket{\bol\psi_0}$ is,
\begin{equation}\label{51}
\hat{\mathcal{U}}_{diss}\ket{0}\ket{\bol\psi_0}=\ket{0}\hat{K}_0\ket{\bol\psi_0}+\frac{1}{2}\ket{1}(\hat{K}_{0z}-\hat{K}_{0z}^\dagger)\ket{\bol\psi_0}.
\end{equation}
Again, a measurement on the first qubit provides the desired result. The quantum circuit representation for simulation of 
the Suzuki-Trotter dynamics \eqref{24}, taking into consideration Eqs.\eqref{50} and \eqref{51}, is depicted in Fig.\ref{fig:2}.
The probability of measuring the $0$-bit value qubit in the output state \eqref{51} is $p_0=\mel{\bol\psi_0}{\hat{K}_0^2}{\bol\psi_0}$.  This obeys 
the same bounds (\eqref{new2}, \eqref{new3}) and subsequently pertain to analogous considerations as in Sec.\ref{sec:3.1.2}.

The remaining question is whether the implementation cost of $\hat{U}_{select}$ scales favorably compared to that of $\hat{\mathcal{U}}_{diss}$ 
from the previous section. Given the definition in \eqref{49}, $\hat{U}_{select}$ is a $2d\times 2d$ diagonal operator,
\begin{equation}
\hat{U}_{select}=\begin{bmatrix}
\hat{K}_{0z}&0\\
0&\hat{K}_{0z}^\dagger
\end{bmatrix},
\end{equation}
which contains $r$ two-level $z$-rotations $\hat{\mathcal{R}}_z(\theta_l)$ compared to the $r$ two-level $y$-rotations 
$\hat{\mathcal{R}}_y(\theta_l)$  of Eq.\eqref{37}. As a result, the diagonal nature of $\hat{U}_{select}$ allows for an 
implementation in $2^n(1-\frac{1}{1+2N})-3$ alternating CNOTs and single-qubit $\hat{R}_z(\theta_l)$ rotations \cite{Bullock}. 
Thus, to leading order, a quadratic improvement is achieved as compared to the scaling of the physical dilation in \eqref{38}. 
The LCU method produces the same dilation method, specialized for diagonal operators, as in \cite{Schlimgen}.

\subsection{Case study: weakly dissipative homogeneous dielectric medium}\label{sec:4.1}
When the medium is a simple, $N=1$, homogeneous dielectric, $\Omega_e(\bol r),\omega_{e}(\bol r),\gamma_e(\bol r)=constants$ and $\Omega_m(\bol r),\omega_{m}(\bol r),\gamma_m(\bol r)=0$, the diagonal matrix  $\hat{U}_{select}$ contains repetitive values. It is then possible to further reduce the implementation cost of $\hat{U}_{select}$ to polynomial scaling $\textit{O}[poly(n)]$, \cite{Hogg}. Similarly, according to the discussion in Sec.\ref{sec:2.1}, a QLA implementation of the unitary part $e^{-i\delta t\hat D}$ for the homogeneous case scales as $\textit{O}(n^2)$. Consequently, the overall implementation scaling of the LCU quantum circuit of Fig.\ref{fig:2} is $\textit{O}[poly(n)+n^2]$. For a quantum simulation of total time $t_{total}=N_t\delta t$ the overall number of gates, taking into consideration the the local post-selection complexity of Sec.\ref{sec:3.1.2}, reads $\textit{O}\big[\frac{N_t}{p_0^{N_t}}(poly(n)+n^2)\big]$.

For a weakly dissipative medium in a specific frequency window \cite{Landau} applies that,
\begin{equation}\label{time}
\lambda_{max}>>\gamma,\quad \Delta t_{uni}<<\Delta t_{diss}.
\end{equation}
Hence, the dissipative time scale $\Delta t_{diss}$ corresponds to the large scale temporal dynamics. Therefore, selecting a Trotter time scale $\delta t\sim\Delta t_{uni}$  a high success local post-selection process, Eq.\eqref{new4}, is guaranteed. The number $N_t$ of the required Trotter steps to obtain the desired approximation within an error $\varepsilon$ is provided in Eq.\eqref{new6},
\begin{equation}\label{number of steps}
N_t=\frac{2\lambda_{max}t_{total}}{W(2\varepsilon/\gamma t_{total})},
\end{equation}
where $W(x)$ is the Lambert W function. The scaling law Eq.\eqref{number of steps} for the number of Trotter steps $N_t$ is interwoven with the different physical time-scales. For instance, aiming to simulate the dynamics for long total time $t_{total}\sim\kappa\Delta t_{diss}$ with $\kappa>>1$ with constant error $\varepsilon\sim\textit{O}(1)$ we obtain from Eq.\eqref{number of steps},
\begin{equation}\label{long time}
N^l_t\sim\kappa^2\frac{\lambda_{max}}{\gamma}.
\end{equation}
For simulation time $t_{total}=\kappa\Delta t_{uni}<\Delta t_{diss}$ with $\kappa>1$, reflecting the transient dynamics, the number of required Trotter steps for constant error is,
\begin{equation}\label{transient}
N^t_t\sim\frac{2\kappa}{W(2\lambda_{max}/\kappa\gamma)}.
\end{equation}
Finally, for short-scale simulation time $t_{total}=\kappa\Delta t_{uni}$  with $\kappa\sim\textit{O}(1)$,
\begin{equation}\label{short time}
N^s_t\sim\frac{2\kappa}{\ln{(2\lambda_{max}/\kappa\gamma)}}.
\end{equation}
In derivation of Eqs.\eqref{long time}-\eqref{short time} the  following approximations $W(x<<1)\approx x$, $W(x>>1)\approx\ln{x}$ have been used.

Consequently, the total number of gates $N_{gates}$, including the post-selection complexity, required for simulation of the various temporal regimes with constant error $\varepsilon$ is,
\begin{equation}\label{finalgates}
N_{gates}=\begin{cases}
    \textit{O}\bigg[\kappa^2\frac{\lambda_{max}}{\gamma p_0^{N^s_t}}(poly(n)+n^2)\bigg],\quad \text{for} \quad \kappa>>1,\\\\
    \textit{O}\bigg[\frac{2\kappa}{p_0^{N^l_t}W(2\lambda_{max}/\kappa\gamma)}(poly(n)+n^2)\bigg],\quad \text{for}\quad \kappa>1,\\\\
    \textit{O}\bigg[\frac{2\kappa}{p_0^{N^s_t}\ln(2\lambda_{max}/\kappa\gamma)}(poly(n)+n^2)\bigg],\quad \text{for}\quad \kappa\sim\textit{O}(1),
\end{cases}
\end{equation}
where $p_0^{N^m_t}=\mel{\bol\psi_0}{\hat{K}_0^{2N_t^m}}{\bol\psi_0}$, for $m=s,t,l$.
and $\lambda_{max}=\lambda_{max}(n,\Omega_{e},\omega_{e})$. Evidently, Eq.\eqref{finalgates} is in accordance with the  physical constrains and the transient nature of evolution, features that cannot be captured through quantum imaginary time evolution approaches \cite{Motta, Kamakari} which would be more suitable for calculation of low frequency modes rather than physical evolution implementation. In the same manner, allowing for different and inhomogeneous dissipation rates will result into  complex temporal dynamics with different local basins of attraction, rendering the task of finding the correct solution very difficult even for modern variational quantum algorithms \cite{Watad}.

\section{Conclusions}\label{sec:5}
An effective simulation of dynamics in dissipative classical systems is inherently challenging 
for quantum computers due to the loss of unitary evolution. In this paper, we have focused on dissipative and 
dispersive electromagnetic media in which energy loss appears as an anti-Hermitian, diagonal part in the Schr\"odinger representation of Maxwell equations. Using the Suzuki-Trotter approximation, we untangle the unitary evolution from the non-unitary, enabling us to concentrate exclusively on the non-unitary evolution.

The interconnection between the dissipation in dispersive electromagnetic media and the dissipation in open quantum systems serves as a first step in a dilation process which relies on an augmented Kraus representation \eqref{31} and the direct sum structure of the dissipative $r$-dimensional subspace. In this way, we only need one ancillary qubit to model the  environment, in contrast to the respective formulation of non-unitary 
quantum evolution through quantum channels \cite{Hubisz}. This physical dilation requires $\textit{O}(2^{n-1}n^2)$ CNOTs 
and one qubit $y$-rotations. The second algorithm does not use the physical connection between classical and quantum processes but implements the non-unitary evolution through the LCU Lemma in $\textit{O}(2^n)$ CNOTs and one qubit $z$-rotations. However, while the second method is quadratically better, it lacks the physical interpretation which is essential when the anti-Hermitian 
part in the Hamiltonian does not correspond to pure dissipation but also accommodates amplification (gain) effects. In such cases, a physical dilation based on the construction of the proper quantum channel for the classical dynamics may be advantageous in terms of physical information that can be extracted from the quantum implementation, compared to the LCU dilation. This is evident in Sec.\ref{sec:3.3} where the PTP NINO quantum operations are considered as well as in the PTP dilation method of \cite{Schlimgen2}.

Combining the results from Secs.\ref{sec:3}, \ref{sec:4} and those in reference \cite{Koukoutsis} in conjunction with a quantum simulation implementation  of Maxwell equations for the lossless case, (possibly with QLA) we can realize a full wave simulation of electromagnetic wave propagation and scattering in complex media, such as magnetized plasmas and meta-materials, in a quantum computer. For such situations, we show that employing the LCU method for the case of a homogeneous medium with a single dissipation rate the implementation cost is reduced to $\textit{O}[poly(n)]$ and the respective circuit depth of the total simulation is strongly correlated with the physical time-scales. As a result, we open a path for future research for quantum computing to make an impact in the field of computational electromagnetism for applications. Our set up, Eqs.\eqref{6} and \eqref{5}, considers minimal assumptions while respecting Kramer-Kroning  causality relations \cite{Landau} in the whole frequency spectrum and extending beyond $\mathcal{PT}$-symmetric models \cite{Zyablovsky}.

Finally, it is important to note that, when computationally feasible, both methods can be applied to other classical dissipative systems by diagonalizing the Hermitian dissipative operator $\hat{D}_{diss}$ in the respective Schr\"odinger equation \eqref{23} as $\hat{D}_{diss}=\hat{V}\hat{\Delta}\hat{V}^\dagger$. 
Then the diagonal operator $\hat{\Delta}$ leads to a diagonal non-unitary part as in Eq.\eqref{24}, which can be implemented using the techniques described in this paper and in Ref.\cite{Schlimgen}. The implementation cost of unitary operators $\hat{V},\hat{V}^\dagger$ is directly related to the dimension of the dissipative subspace.

\section*{Acknowledgments}

This work has been carried out within the framework of the EUROfusion Consortium, funded by the European Union via the Euratom Research and Training Programme (Grant Agreement No 101052200 — EUROfusion). Views and opinions expressed are however those of the authors only and do not necessarily reflect those of the European Union or the European Commission. Neither the European Union nor the European Commission can be held responsible for them.
A.K.R is supported by the Department of Energy under Grant Nos. DE-SC0021647 and DE-FG02-91ER-54109.
G.V is supported by the Department of Energy under Grant No. DE-SC0021651.







\end{document}